\documentclass{article}

    \usepackage[preprint]{neurips_2024}

\usepackage[utf8]{inputenc} 
\usepackage[T1]{fontenc}    
\usepackage{hyperref}       
\usepackage{url}            
\usepackage{booktabs}       
\usepackage{amsfonts}       
\usepackage{nicefrac}       
\usepackage{microtype}      
\usepackage{xcolor}         

\usepackage{graphicx,subfigure,multirow}
\usepackage{bm}
\usepackage{amsmath}
\usepackage{booktabs}
\usepackage{enumitem}
\usepackage{bbding}
\usepackage{color}
\usepackage{natbib}
\setcitestyle{numbers,square}
\usepackage{diagbox}
\usepackage{wrapfig}
\usepackage{booktabs}
\usepackage{caption}
\usepackage{pifont}
\usepackage{makecell}

\hypersetup{hidelinks}

\newcommand{\M}{MelodyLM}

\title{Accompanied Singing Voice Synthesis with Fully Text-controlled Melody}

\author{%
  Ruiqi Li$^1$, Zhiqing Hong$^1$, Yongqi Wang$^1$, Lichao Zhang$^1$, Rongjie Huang$^1$, \\ \textbf{Siqi Zheng$^2$, Zhou Zhao$^1$} \\
  $^1$Zhejiang University \& $^2$Alibaba Group \\
  \texttt{ruiqili@zju.edu.cn} \\
}

\begin{document}

\maketitle

\begin{abstract}

Text-to-song (TTSong) is a music generation task that synthesizes accompanied singing voices. Current TTSong methods, inherited from singing voice synthesis (SVS), require melody-related information that can sometimes be impractical, such as music scores or MIDI sequences. We present MelodyLM, the first TTSong model that generates high-quality song pieces with fully text-controlled melodies, achieving minimal user requirements and maximum control flexibility. MelodyLM explicitly models MIDI as the intermediate melody-related feature and sequentially generates vocal tracks in a language model manner, conditioned on textual and vocal prompts. The accompaniment music is subsequently synthesized by a latent diffusion model with hybrid conditioning for temporal alignment. With minimal requirements, users only need to input lyrics and a reference voice to synthesize a song sample. For full control, just input textual prompts or even directly input MIDI. Experimental results indicate that MelodyLM achieves superior performance in terms of both objective and subjective metrics. Audio samples are available at \href{https://melodylm666.github.io}{https://melodylm666.github.io}. 

\end{abstract}

\section{Introduction}

    \begin{figure*}[h]
        \centering
        \includegraphics[width=\textwidth]{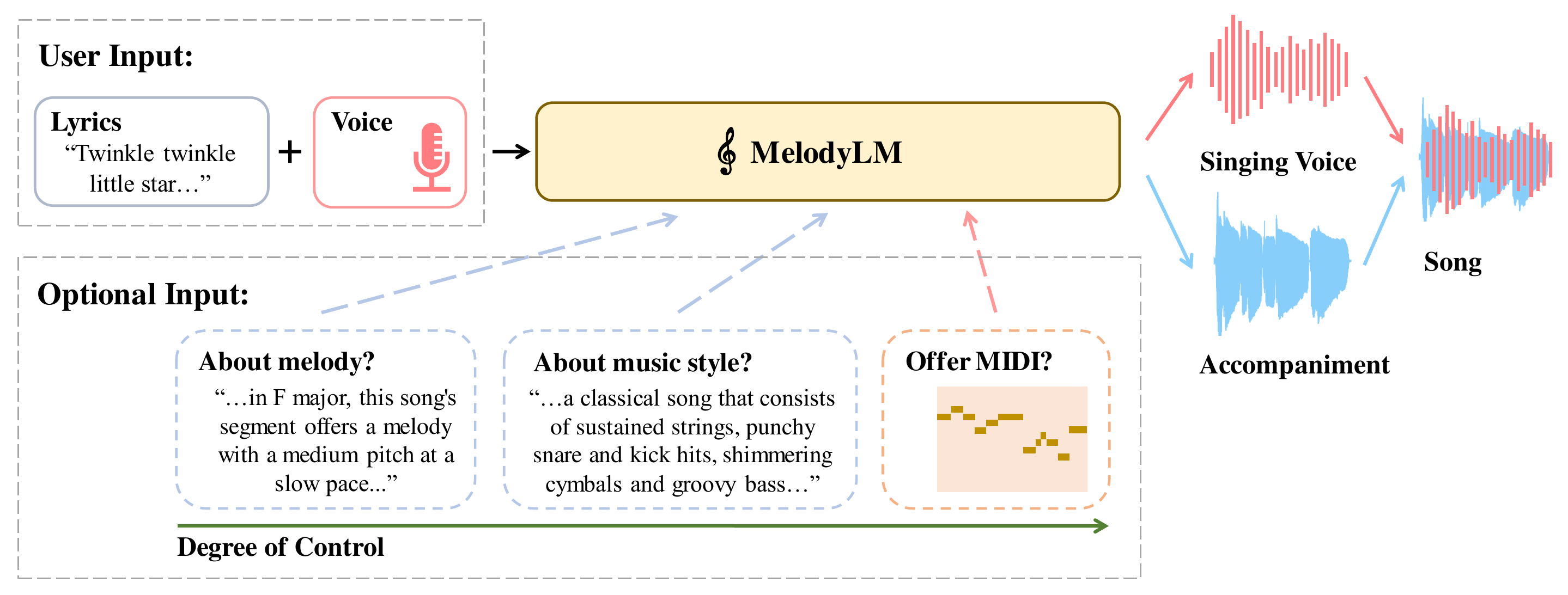}
        \setlength{\abovecaptionskip}{-0.3cm}
        \setlength{\belowcaptionskip}{-0.4cm}
        \caption{
            Overview of \M.  
        }
        \label{fig:main-model}
    \end{figure*}

Text-to-song (TTSong), or accompanied singing voice synthesis (ASVS), refers to generating song samples given natural language descriptions. In this task, a song piece is defined as a musical composition that incorporates both a vocal track and instrumental accompaniments \cite{zhiqing2024text}. Different from music generation \cite{dong2018musegan, agostinelli2023musiclm, huang2023noise2music, copet2024simple}, song generation requires meaningful and perceptible vocals. In contrast to singing voice synthesis (SVS) \cite{Liu_Li_Ren_Chen_Zhao_2022, he2023rmssinger, zhang2024stylesinger}, TTSong also necessitates generating high-quality and long-distance coherent accompaniments.

Previous works \cite{zhiqing2024text} divide TTSong into two stages: 1) vocal synthesis given music score
(or MIDI) and lyrics; and 2) accompaniment generation, conditioned on the vocal signals from stage 1. However, asking users to provide music scores is often unrealistic in practice, as it requires a certain level of musical expertise. Currently, research on TTSong with fully text-controlled melody is still lacking to our knowledge, facing challenges such as data scarcity and complexity.

Recent advancements in symbolic music generation \cite{dong2018musegan, yu2022museformer, ding2024songcomposer} and language models (LMs) \cite{devlin2018bert, brown2020language, wei2021finetuned, touvron2023llama} provide new insights into addressing this problem. A natural idea is to introduce an additional stage 0 before stage 1, where the melody-related features, like music scores, are generated from text descriptions in advance. The development of audio language models \cite{borsos2023audiolm, wang2023neural, yang2023uniaudio} also provides the foundation for unifying the first two stages within the framework of sequence modeling. 

In this paper, we introduce \M, the first TTSong model that generates high-quality song pieces with fully text-controlled melodies. We propose a 3-stage framework, namely, text-to-MIDI, text-to-vocal, and vocal-to-accompaniment, for coherent song generations and solving the problem of data scarcity. The first two stages can be unified within a language modeling framework, and the third stage is implemented based on a latent diffusion model (LDM) \cite{rombach2022high, huang2023make, liu2023audioldm, schneider2023mo} with hybrid conditioning for a better alignment with vocals. 
To improve the controllability of melodies, we incorporate automatically transcribed MIDI sequences as the intermediate melody-related features, and construct corresponding textual prompts containing musical attributes such as key, tempo, etc. 
The MIDI note events are considered discrete tokens and are modeled using a multi-scale transformer \cite{yang2023uniaudio}, which is also used for the vocal codec language model in stage 2. For controllable accompaniment generation, we leverage a music captioning model \cite{doh2023lp} to generate pseudo descriptions as accompaniment-related prompts. Random drop is applied for all text prompts except lyrics. Therefore, in our settings, the minimal requirement of user input is lyrics and a vocal reference prompt, where the latter could be further reduced if the vocal prompts are stored in advance. 
For a fully controllable generation, users can input prompts or even MIDI sequences.

Experiments on a combination of open-source and web-crawled Mandarin pop song corpora show that \M~has the capability of generating accompanied singing voices given pure textual descriptions and vocal prompts. 
The main contributions of this work are summarized as follows:
\begin{itemize}[leftmargin=*]
    \item We propose \M, the first accompanied singing voice generation model that takes pure textual descriptions and vocal references as conditions, eliminating the need for users to input music scores.
    \item We design a MIDI language model and a vocal codec language model under the same multi-scale framework, which explicitly models MIDI as the intermediate melody-related features and provides considerable controllability.
    \item We present a latent diffusion model with hybrid conditioning for vocal-to-accompaniment, achieving controllable high-quality accompaniment generation.
    \item Experimental results of objective and subjective evaluations reveal the effectiveness of the hierarchical design, with a subjective rating of 75.6/100 for \M~against 79.8 for the best baseline. 
\end{itemize}

\section{Related Work}

\subsection{Singing Voice Synthesis}

SVS aims to generate singing voice signals conditioned on lyrics and music scores (or F0 contours). DiffSinger \cite{Liu_Li_Ren_Chen_Zhao_2022} designs a shallow diffusion mechanism to address the over-smooth issues in the general text-to-speech (TTS) field. To facilitate style transfer and zero-shot synthesis, NaturalSpeech 2 \cite{shen2023naturalspeech} and StyleSinger \cite{zhang2024stylesinger} leverage a reference voice clip for timbre and style extraction. To bridge the gap between realistic music scores and fine-grained MIDI annotations, RMSSinger \cite{he2023rmssinger} proposes word-level modeling with a diffusion-based pitch prediction approach. MIDI-Voice \cite{byun2024midi} incorporate MIDI-based prior for expressive zero-shot generation. Open-source singing voice corpora also boost the development \cite{huang2021multi, zhang2022m4singer, wang2022opencpop}, although their quantity is still much smaller compared to speech datasets. PromptSinger \cite{wang2024prompt} is the first to attempt guiding singing voice generation using text descriptions, which focuses more on speaker identity and timbre control. Melodist \cite{zhiqing2024text} is the first TTSong model, which adopts two autoregressive transformers to sequentially generate vocal codec tokens and accompaniment codec tokens. 

\subsection{Music Generation}

Music generation is a general task that includes symbolic music generation, waveform generation, and accompaniment generation. MuseGAN \cite{dong2018musegan} achieves symbolic music generation via a GAN-based approach. 
SongMASS \cite{sheng2021songmass} designs a songwriting method that generates lyrics or melodies, conditioned on each other. SongComposer \cite{ding2024songcomposer} proposes a music large language model (LLM) for song composition, which can compose melodies and lyrics with symbolic song representations. 

Inspired by the two-stage modeling in audio generation \cite{borsos2023audiolm}, MusicLM \cite{agostinelli2023musiclm} adopts a cascade of transformer decoders to sequentially generate the semantic tokens and the acoustic tokens, conditioned on joint textual-music representations from MuLan \cite{huang2022mulan}. MusicGen \cite{copet2024simple} proposes a novel codebook interleaving patterns to generate music codec tokens in a single transformer decoder, while MeLoDy \cite{lam2024efficient} introduces an LM-guided diffusion model that efficiently generates music audios. MusicLDM \cite{chen2024musicldm} incorporates beat-tracking information and applies data augmentation through latent mixup to address the potential plagiarism issue in music generation. However, music generation models are not required to produce perceptible singing voices, which is still a challenge. 

In addition, several works focus on singing-to-accompaniment generation. SingSong \cite{donahue2023singsong} is proposed to generate instrumental music to accompany input vocals. Melodist \cite{zhiqing2024text} incorporates a transformer decoder to achieve controllable accompaniment generation.

\section{Method}

This section introduces \M. To address the issue of limited data crawled from the web, we divided this task into three hierarchical stages to explicitly model information flows. Stage 0 serves as a MIDI language model (MIDI-LM) over text prompts and discrete MIDI note events. For stage 1, we adopt a widely-used vocal acoustic language model (Vocal-LM) \cite{yang2023uniaudio} to implicitly align the lyrics with MIDI tokens and generate quantized acoustic units from a Soundstream audio tokenizer \cite{zeghidour2021soundstream}. In stage 2, an LDM is adopted for the high-quality reconstruction of accompaniments, where we leverage a hybrid conditioning mechanism to balance both textual prompting and vocal conditioning. The overall model design is illustrated in \autoref{fig:main-model}.

\subsection{Information Flow}

    \begin{figure*}[t]
        \centering
        \includegraphics[width=\textwidth]{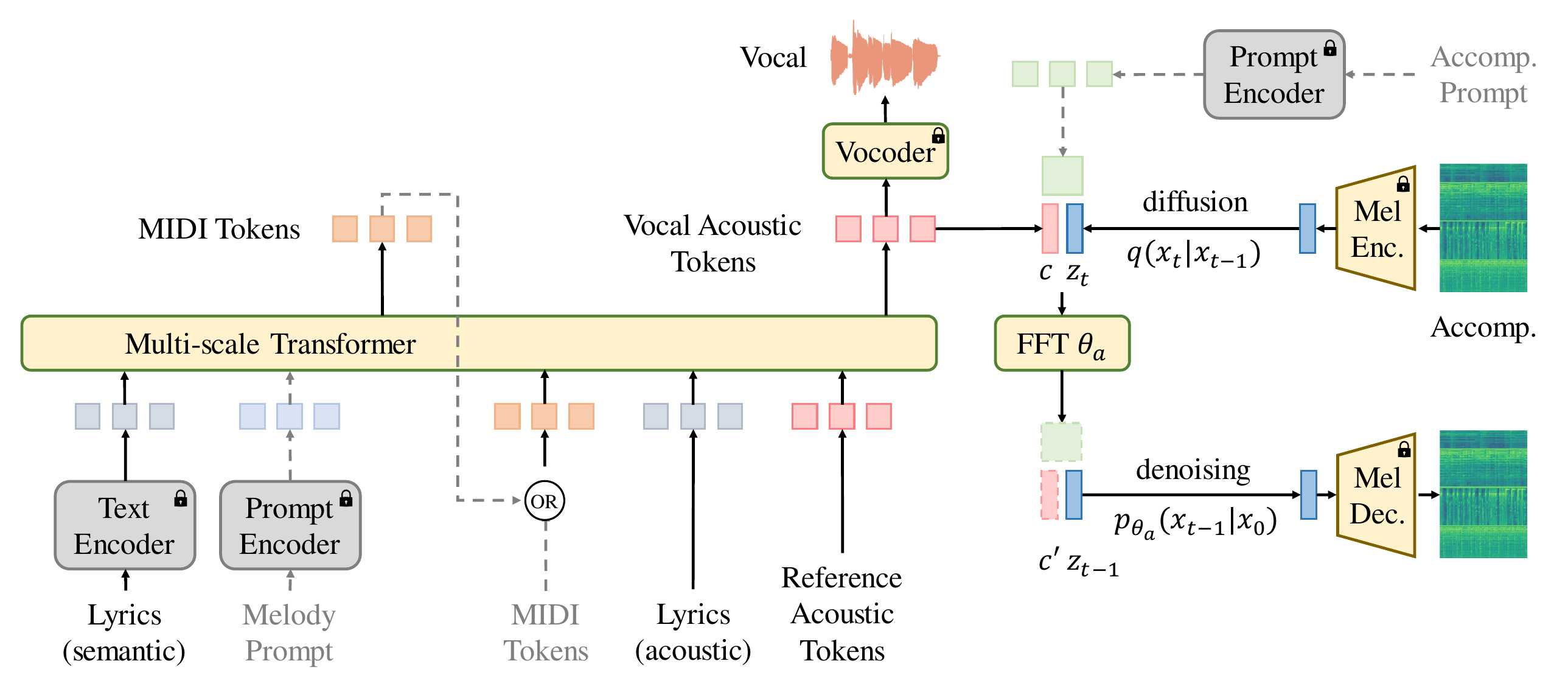}
        \setlength{\abovecaptionskip}{-0.3cm}
        \setlength{\belowcaptionskip}{-0.5cm}
        \caption{
            The overall architecture. The gray dashed lines indicate optional inputs. Modules printed with a \textit{lock} are frozen during the training stage. Lyrics (semantic) and (acoustic) are essentially the same input lyrics, except the former provides potential semantic information (hence processed by a text encoder) and the latter only provides pronunciation-related acoustic information. 
        }
        \label{fig:main-model}
    \end{figure*}

In this section, we introduce the key information and its flow, especially the extraction, construction, and utilization of important intermediate features. 

\subsubsection{MIDI}

The choice of the intermediate melody-related feature is tricky. The fundamental frequency (F0) of vocal is a commonly used feature in SVS \cite{Liu_Li_Ren_Chen_Zhao_2022}. However, the F0 contour is strongly correlated with prosody, making it difficult to decouple from other voice attributes (like energy) or even the singer's identity \cite{ren2022prosospeech}. In the meanwhile, the chromagram is also used as a disentangled melody condition \cite{copet2024simple}, but it only contains 12 bins (one octave), which is not sufficient to fully represent the melody.

MIDI satisfies all the requirements. MIDI can be maximally decoupled from the singer's identity information, and a range of 32 (G\#1, 51.9Hz) to 80 (G\#5, 830.6Hz) covers most vocal melodies. Since there is no open-source large song corpus with available MIDI annotations, we use ROSVOT \cite{li2024robust} to extract the MIDI sequences from the demixed vocal tracks. A MIDI sequence $\boldsymbol{m}$ consists of $L_m$ note events, where each note event $m^l$ has two attributes, pitch $m_p^l$, and duration $m_d^l$: $\boldsymbol{m} = [(m_p^1, m_d^1), (m_p^2, m_d^2), ..., (m_p^{L_m}, m_d^{L_m})]$, $l = 1, ..., L_m$, where the durations are quantized into number of frames. This intrinsic discrete attribute makes MIDI naturally suitable for language models. We leverage the MIDI sequences as a bridge to connect the text prompts and the target singing voices. 

\subsubsection{Textual Prompt}

\paragraph{Melody-related Prompt}
For controllable melody generation, we construct artificial textual prompts to deliver melody-related information. We extract musical attributes like key, tempo, average pitch, etc., from the waveforms and the extracted MIDI sequences. We produce $N_{\text{melody}}$ textual templates for musical attributes to construct melody-related natural language prompts. Details are listed in \autoref{appendix:prompt}


\paragraph{Accompaniment-related Prompt}
We adopt LP-MusicCaps \cite{doh2023lp}, a music captioning model, on the original song piece $\boldsymbol{y}$ to generate pseudo-captions regarding global timbre perception, sound quality, instrument usage, musical style, and overall emotion. This prompt provides accompaniment-related information for music reconstruction.

\subsubsection{Voice Tokenization}

Inspired by previous works \cite{yang2023uniaudio, wang2024prompt}, we use Soundstream, an audio tokenizer, to produce discrete acoustic units. The tokenizer adopts residual vector quantization with $N_q = 8$ codebooks with size $K$ and produce a vector sequence $\boldsymbol{A} = [\boldsymbol{a}_1, \boldsymbol{a}_2, ..., \boldsymbol{a}_N]$ given a vocal signal $\boldsymbol{y}_{\text{vocal}}$, where $\boldsymbol{a}_i = [a_t^1, a_t^2, ..., a_t^{N_q}]$, being a vector consists of $N_q$ codes at timestep $i \in [1, N]$, and $a_i^{\tau} \in [0, K-1]$, $\forall \tau \in [1, N_q]$. We choose the tokens from the first 3 codebooks in the subsequent multi-scale language modeling, lightening the burden of autoregressive generation. A HiFi-GAN vocoder \cite{kong2020hifi} is adopted for high-quality audio reconstruction. 

    \begin{figure*}[t]
        \centering
        \includegraphics[width=\textwidth]{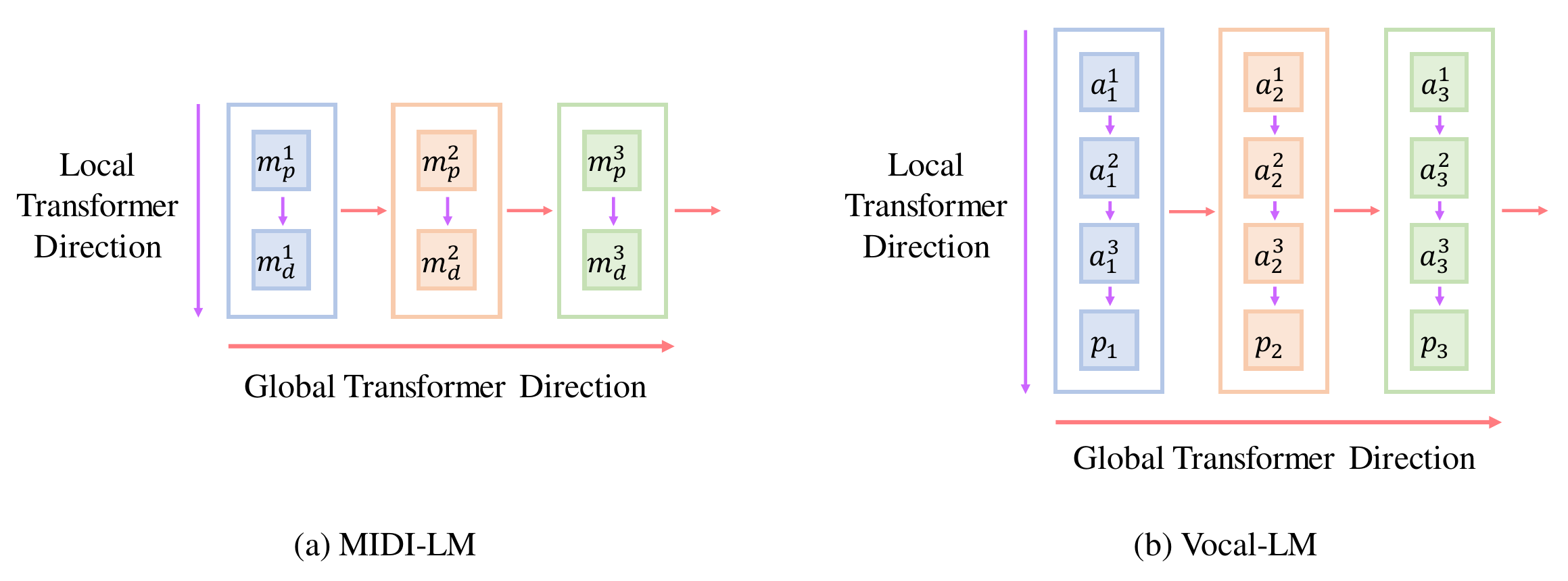}
        \setlength{\abovecaptionskip}{-0.1cm}
        \setlength{\belowcaptionskip}{-0.5cm}
        \caption{
            Multi-scale language modeling for MIDI tokens and vocal acoustic tokens. 
        }
        \label{fig:lm}
    \end{figure*}

\subsection{MIDI and Vocal Language Modeling}

The discrete representations of MIDI and vocal acoustic units share a common characteristic: there are multiple tokens at each timestep. The $l$th MIDI event $m^l$ contains two tokens, pitch $m_p^l$ and duration $m_d^l$, while there are $N_q$ parallel vocal acoustic tokens from different codebooks at each timestep $i$. Inspired by \cite{yang2023uniaudio}, we design a multi-scale transformer to autoregressively model this hierarchical distribution. Without loss of generality, given a sequence $X = [\boldsymbol{x}_1, \boldsymbol{x}_2, ..., \boldsymbol{x}_N]$, where each element at timestep $i = 1, 2, ..., N$ consists of $P$ parallel tokens $\boldsymbol{x}_i = [x_i^1, x_i^2, ..., x_i^P]$, we concatenate these $P$ tokens along the channel axis (denoted as $\oplus$) and adopt a global transformer $\theta_{\text{G}}$ to model the temporal distribution:
\begin{equation}
    p(\boldsymbol{o}|\boldsymbol{h}; \theta_{\text{G}}) = \prod_{i=1}^N p(o_i|\boldsymbol{o}_{<i}, \boldsymbol{h}_{<i}; \theta_{\text{G}}), \quad \text{where } h_i = x_i^1 \oplus x_i^2 \oplus ... \oplus x_i^P
\end{equation}
$o_i$ represents the global context information at $i$. After that, we stack all $N$ elements and propagate along the internal direction using a local transformer $\theta_L$, conditioned on the global context $\boldsymbol{o}$:
\begin{equation}
    p(\boldsymbol{x}_i|o_i; \theta_{\text{L}}) = \prod_{\tau=1}^{P} p(x_i^{\tau}|\boldsymbol{x}_i^{<\tau}, z_i; \theta_{\text{L}})
\end{equation}
This global context conditioning is implemented by simply projecting $o_i$ to the same size as $x_i^{\tau}$ and performing an element-wise summation. During the inference stage, the two transformers run alternatively: the global model $\theta_{\text{G}}$ generates the context vector $o_i$ at timestep $i$, and the local model $\theta_{\text{L}}$ generates $P$ tokens conditioned on $o_i$ autoregressively. 

We applied this framework to the modeling of both MIDI and vocal acoustic tokens, where the $P$ value for MIDI is 2 and vocal is 3. It is worth mentioning that the text representations have to repeat themselves for $P$ times at each timestep to fit into the framework. During the training stage, only the MIDI sequences and vocal sequences after prompts are considered in the loss. For MIDI modeling, we use text prompts to directly predict target note events, $m_p^l$ and $m_d^l$. 

For vocal modeling, we expand the input MIDI pitches along the time axis according to the corresponding durations: $\overline{\boldsymbol{m}} = [m_p^{1, 1}, m_p^{1, 2}, ..., m_p^{1, m_d^1}, ..., m_p^{L_m, 1}, m_p^{L_m, 2}, ..., m_p^{L_m, m_d^{L_m}}]$, so that $\overline{\boldsymbol{m}}$ now has the same temporal length as the target vocal codec sequence. This is because we find that LMs are more adept at perceiving positional information rather than numerical duration information. Inspired by \cite{li2023alignsts}, this structure forms an implicit re-alignment between the lyrics and the target melody, where the latter queries the possible word to pronounce. In addition, for better prosody modeling, we add a 4th token $p_t$ at each timestep to predict the target F0 contour, as shown in \autoref{fig:lm}. 

\subsection{Accompaniment Generation with Hybrid Conditioning}

    \begin{figure*}[t]
        \centering
        \includegraphics[width=0.9\textwidth]{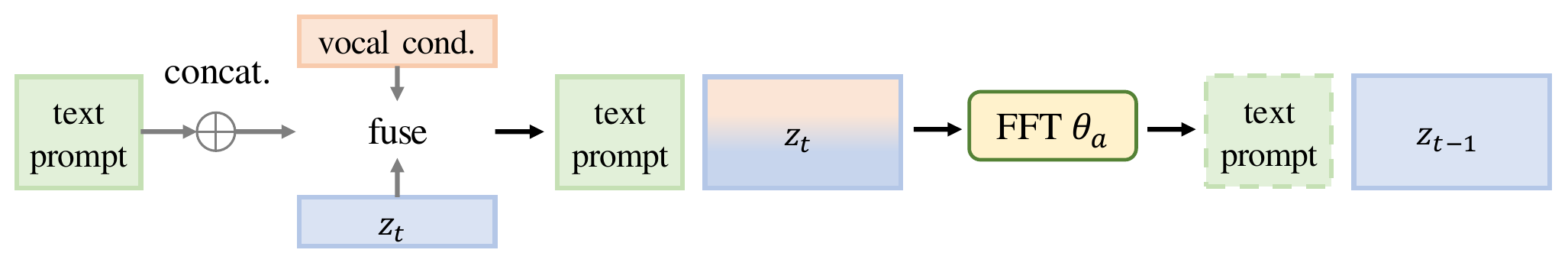}
        \setlength{\abovecaptionskip}{0.1cm}
        \setlength{\belowcaptionskip}{-0.5cm}
        \caption{
            Accompaniment latent diffusion with hybrid conditioning. 
        }
        \label{fig:ldm}
    \end{figure*}

A latent diffusion model is adopted for the accompaniment generation, since the target temporal length is acquired after vocal generation. To facilitate both in-context learning from textual prompts and vocal conditioning, we choose a multi-layer feed-forward transformer (FFT) \cite{peebles2023scalable} as the denoiser model. To achieve accurate temporal alignment, a hybrid conditioning mechanism is designed.
\paragraph{Forward Process}
Given a target latent feature $z_0$ from a pre-trained VAE model \cite{kingma2013auto}, which can be sampled from a distribution $q(z_0)$, the forward diffusion process is a Markov chain that gradually perturbs $z_0$ to pure noise \cite{ho2020denoising}. At each diffusion step $t \in [1, T]$, a tiny Gaussian noise is added to $z_{t-1}$ to obtain $z_t$. 
\paragraph{Reverse Process}
The reverse process is a Markov chain with learnable parameters $\theta_a$ from $z_T$ to $z_0$. Given the conditions: a sequence of textual representations $\boldsymbol{s} \in \mathbb{R}^{n \times d}$ and vocal acoustic units $\boldsymbol{a} \in \mathbb{R}^{N \times d}$, we approximate the Gaussian distribution at each timestep $t$ by the denoising model $\theta_a$:
\begin{equation}
    p_{\theta_a}(z_{t-1}|z_{t}, \boldsymbol{s}, \boldsymbol{a}) := \mathcal{N}(z_{t-1}; \mu_{\theta_a}(z_t, t, \boldsymbol{s}, \boldsymbol{a}), \sigma^2 \mathbf{I}).
\end{equation}
where we implement the hybrid conditioning through channel-wise vocal injection and the partial noising \cite{ho2020denoising} strategy. Specifically, we concatenate the vocal acoustic tokens $\boldsymbol{a}$ and the corrupted sample $z_t$ along the channel dimension $d$, and use linear projection $W \in \mathbb{R}^{2d \times d}$ to obtain a fusion feature $\widetilde{z}_{t} = (\boldsymbol{a} \oplus z_t) W$, $\widetilde{z}_{t} \in \mathbb{R}^{N \times d}$, which is further concatenated with the textual prompts $\boldsymbol{s}$ along the temporal dimension (denoted as $\text{concat}(\cdot)$) to acquire the expanded feature: $Z_t = \text{concat}(\boldsymbol{s}, \widetilde{z}_{t})$, $Z_t \in \mathbb{R}^{n+N}$, as shown in \autoref{fig:ldm}. Note that the fraction corresponding to the textual prompts in the recovered $Z_0$ is dropped, and the loss is only computed concerning $z_0$, which forms an implicit cross-attention operation. We believe the channel-wise fusion of vocal information provides fine-grained temporal guidance for accurate reconstruction, and the prompting strategy serves as an in-context learning mechanism for controllable generation. 


\subsection{Training and Inference}

The three stages are trained separately, and the detailed procedure is listed in \autoref{appendix:train}. For inference, as stated before, \M~is designed for both user requirement reduction and full controllability. Therefore, the only requirement is the lyrics and the reference vocal. If a user wishes to provide the target MIDI sequence to precisely control the melody, we dismiss stage 0. Otherwise, we use the MIDI-LM to generate a target melody sequence, conditioned on the lyrics and the optional prompts. The target vocal tokens are generated by the Vocal-LM, conditioned on the MIDI sequence, the lyrics, and the reference vocal tokens. Finally, the LDM and the VAE model generate the accompaniment, conditioned on the vocal tokens and the optional prompts. 


\section{Experimental Setup}

\subsection{Data}

We use a combination of open-source and web-crawled Mandarin pop song corpora to train the first two stages. The open-source corpus is vocal-only, including Opencpop \cite{wang2022opencpop}, M4singer \cite{zhang2022m4singer}, and OpenSinger \cite{huang2021multi}. Since there are no available open-source datasets with available accompaniments, we crawl about 17K Mandarin songs from YouTube and a well-known music website with 237 artists, following \cite{zhiqing2024text}. After preprocessing, the combined dataset yields around 355K voiced clips with an average length of 12 seconds, totaling approximately 800 hours. For accompaniment generation, we use a filtered fraction of LP-MusicCaps-MSD \cite{doh2023lp}, in addition to the crawled Mandarin pop songs, resulting in a total size of around 1K hours. For evaluation and ablation studies, we leave out 300 in-domain samples each for validation and testing, with no singer overlapping with the training set. 50 of them are chosen from Opencpop and M4Singer for SVS evaluation with GT MIDI input. The details of the datasets and preprocessing procedures are listed in \autoref{appendix:data}. 

\subsection{Implementation and Hyperparameters}

We build a 16-layer global transformer and a 6-layer local transformer for the MIDI-LM, while the Vocal-LM shares a similar architecture but has 20 layers. The sizes of the global transformers are 110M and 320M, respectively. A SoundStream audio tokenizer is used to tokenize 16 kHz vocal waveforms and a unit-based HiFi-GAN is adopted for reconstruction and upsampling to 24 kHz. Mel-spectrograms for accompaniment generation are extracted from 24 kHz waveforms. 
Details of implementation and hyperparameters are listed in \autoref{appendix:imp}.

\subsection{Evaluation}


\paragraph{Objective Metrics}

For MIDI generation, we come up with several metrics to test the controllability. We still use the Krumhansl-Schmuckler algorithm to predict the potential key of the generated MIDI sequences and report the average key accuracy \textbf{KA}\footnote{If the Pearson correlation coefficient of the GT MIDI corresponding to the GT key is $r$, and the predicted MIDI corresponding to the GT key is $\hat{r}$. We define the key accuracy as $\text{KA} = \hat{r}/r$ (only valid if $r \neq 0$).}. We compute the average absolute difference of the average pitches (\textbf{APD}, in semitones) between the GT and the predicted MIDIs, and the average absolute difference of the temporal duration (\textbf{TD}, in seconds). 
In addition, we follow SongMASS \cite{sheng2021songmass} and record the pitch and duration distribution similarity (\textbf{PD} and \textbf{DD}). Melody distance (\textbf{MD}) is also computed through dynamic time warping. 
For the SVS task, we report the F0 frame error (\textbf{FFE}). For the last stage, we compute Frechet audio distance (\textbf{FAD}), Kullback–Leibler divergence (\textbf{KL}), and the CLAP score (\textbf{CLAP}). 

\paragraph{Subjective Metrics}

For the SVS task, we conduct crowd-sourced mean opinion score (MOS) listening tests. Specifically, we score \textbf{MOS-P}, \textbf{MOS-Q}, and \textbf{SMOS} for prosody modeling, overall quality, and singer similarity. For accompaniment generation, we ask the raters to evaluate the audio samples in terms of overall quality (\textbf{OVL}), relevance to the prompt (\textbf{REL}), and alignment with the melody (\textbf{MEL}) of the singing voice. An additional \textbf{OVL-M} score is recorded to evaluate the final mix of the generated vocals and accompaniments. 

\subsection{Baselines}

Since there is no open-source controllable text-to-MIDI model to our knowledge, we construct three variants of the proposed MIDI-LM by implementing different prompt encoders: BERT-large \cite{devlin2018bert}, FLAN-T5-large \cite{chung2024scaling}, and the text encoder of CLAP \cite{elizalde2023clap}. We still compare with SongMASS \cite{sheng2021songmass} for evaluation.
We drop the melody prompt of \M~for uncontrolled generation as another baseline for a fair comparison. Also, SongMASS is designed to generate sheet music scores with structured note durations. Therefore, we provide additional results by rounding off the duration with a granularity of 1/16 note (denoted as "rounded") when computing PD, DD, and MD, following \cite{sheng2021songmass}.

For SVS, we compare our model with 1) DiffSinger \cite{Liu_Li_Ren_Chen_Zhao_2022}, a diffusion-based non-autoregressive generative model; 2) RMSSinger \cite{he2023rmssinger}, an upgraded version of DiffSinger that incorporates natural F0 modeling; and 3) the SVS part of Melodist \cite{zhiqing2024text}. We follow the original design of DiffSinger and RMSSinger, using the global embedding extracted from a voice encoder \cite{resemblyzer} for zero-shot tests. These baselines are all trained with the listed annotated Mandarin datasets, since they require aligned MIDI annotations. For accompaniment generation, we compare the accompaniment part of Melodist, which is re-trained with the same training set as our model. 

\section{Results}


\subsection{MIDI Generation}

\begin{table}[t]
    \caption{Results of MIDI generation.}
    \label{tab:midi-res}
    \footnotesize
    \centering
    \resizebox{0.92\textwidth}{!}{
    \begin{tabular}{lcccccc}
    \toprule
    Methods                     & KA(\%)$\uparrow$  & APD$\downarrow$   & TD$\downarrow$    & PD(\%)$\uparrow$  & DD(\%)$\uparrow$  & MD $\downarrow$  \\ 
    \midrule
    SongMASS                    & 56.6              & 3.84              & 3.02              & 53.1              & 66.3              & \bf3.36              \\
    \M(w/o prompt)              & 54.3              & 3.61              & 5.41              & 53.9              & 26.4              & 4.32              \\
    \M(rounded \& w/o prompt)   & 54.3              & 3.61              & 5.41              & 53.9              & 34.3              & 4.15              \\
    \M(rounded \& w/ FLAN-T5)   & 76.6              & 2.05              & 2.29              & 62.8              & \bf82.4           & 3.54          \\
    \midrule[0.2pt]
    \M(w/ FLAN-T5)              & 76.6              & \bf2.05           & \bf2.29           & \bf62.8           & 40.8              & 3.62              \\
    \M(w/ BERT)                 & \bf77.1           & 2.18              & 2.74              & 60.3              & 39.9              & 3.68              \\
    \M(w/ CLAP)                 & 72.8              & 2.27              & 3.15              & 60.1              & 35.4              & 3.95              \\
    \bottomrule 
    \end{tabular}
    }

\end{table}

\begin{table}[t]
    \caption{Results of accompaniment generation and song production.}
    \label{tab:accomp-res}
    \footnotesize
    \centering
    \setlength\tabcolsep{3pt}
    \resizebox{\textwidth}{!}{
    \begin{tabular}{lccccccccc}
    \toprule
    Methods     & \thead{w/ GT \\ MIDI}     & \thead{w/ GT \\ Vocal}    & FAD$\downarrow$   & KL$\downarrow$    & CLAP$\uparrow$    & OVL$\uparrow$     & REL$\uparrow$     & MEL$\uparrow$     & OVL-M$\uparrow$ \\ 
    \midrule
    Melodist    & -                         & \ding{55}                 & 4.06              & 1.48              & 0.29              & 81.06$\pm$1.88    & 83.10$\pm$1.41    & 70.02$\pm$1.53    & 75.56$\pm$1.34  \\
    Melodist    & -                         & \ding{51}                 & 3.92              & 1.41              & 0.32              & 83.15$\pm$1.69    & 84.61$\pm$1.85    & 71.46$\pm$1.24    & 79.54$\pm$1.47  \\
    \M~(unc.)   & \ding{55}                 & \ding{55}                 & 3.61              & 1.36              & -                 & 83.72$\pm$1.29    & -                 & 70.53$\pm$0.95    & 77.26$\pm$1.06  \\
    \M          & \ding{55}                 & \ding{55}                 & 3.59              & 1.39              & 0.31              & 83.78$\pm$1.35    & 84.71$\pm$1.75    & 70.51$\pm$0.64    & 79.82$\pm$0.96  \\
    \M          & \ding{51}                 & \ding{55}                 & 3.42              & 1.35              & 0.35              & 84.31$\pm$1.67    & 85.95$\pm$1.79    & 72.84$\pm$1.14    & 81.58$\pm$1.25  \\
    \M          & -                         & \ding{51}                 & \bf3.13           & \bf1.31           & \bf0.36           & \bf84.67$\pm$1.23 & \bf86.08$\pm$1.51 & \bf75.19$\pm$0.82 & \bf82.93$\pm$1.04  \\

    \bottomrule 
    \end{tabular}
    }

\end{table}


The results of MIDI generation are listed in \autoref{tab:midi-res}. It can be seen that the duration rounding off operation affects the duration prediction the most, making it somehow an unfair comparison with SongMASS. Also, dropping textual prompts reduces the controllability of \M~dramatically. An interesting finding is that the model has some difficulty perceiving target duration, since an average difference of 2.3 seconds in duration is still nonnegligible, considering an average duration of about 12 seconds for all the clips. The method with a prompt encoder of FLAN-T5 surpasses the other baselines for most of the metrics, but has only a slight advantage over BERT, which is probably because of its larger parameter count and multi-task capability. In terms of the MD metric, \M~underpoerforms SongMASS, but this instead demonstrates that \M~can generate more diverse melodies.

\subsection{Singing Voice Synthesis}

\begin{wraptable}{r}{9cm}
    \caption{Results of SVS.}
    \label{tab:svs-res}
    \footnotesize
    \setlength\tabcolsep{3.5pt}
	\centering
    \resizebox{0.64\textwidth}{!}{
	\begin{tabular}{lcccc}
    \toprule
    Methods             & FEE$\downarrow$   & SMOS$\uparrow$    & MOS-P$\uparrow$   & MOS-Q$\uparrow$   \\
    \midrule
    GT                  & -                 & -                 & 4.11$\pm$0.03     & 4.08$\pm$0.04     \\
    \midrule[0.2pt]
    DiffSinger          & 0.17              & 3.74$\pm$0.11     & 3.66$\pm$0.08     & 3.75$\pm$0.07     \\
    RMSSinger           & 0.09              & 3.76$\pm$0.09     & 3.74$\pm$0.06     & \bf3.79$\pm$0.05  \\
    Melodist            & 0.12              & \bf3.89$\pm$0.06  & 3.79$\pm$0.10     & 3.74$\pm$0.09     \\
    \M~(ours)            & \bf0.08           & 3.81$\pm$0.12     & \bf3.82$\pm$0.08  & 3.76$\pm$0.10     \\
    \bottomrule 
	\end{tabular}
 }
\end{wraptable}

We conduct evaluations of SVS models with GT MIDI input, and the results are listed in \autoref{tab:svs-res}. Due to the contextual learning ability from large-scale data training, \M~can model natural prosody, as reflected in FEE and MOS-P. Specifically, \M~generates voices with various singing techniques, like vibrato or falsetto. A Mel-spectrogram visualization in \autoref{fig:lm} demonstrates that \M~even generates more expressive vibrato than GT. 

\begin{wrapfigure}{r}{0.6\textwidth}
    \vspace{-1em}
    \includegraphics[width=0.6\textwidth]{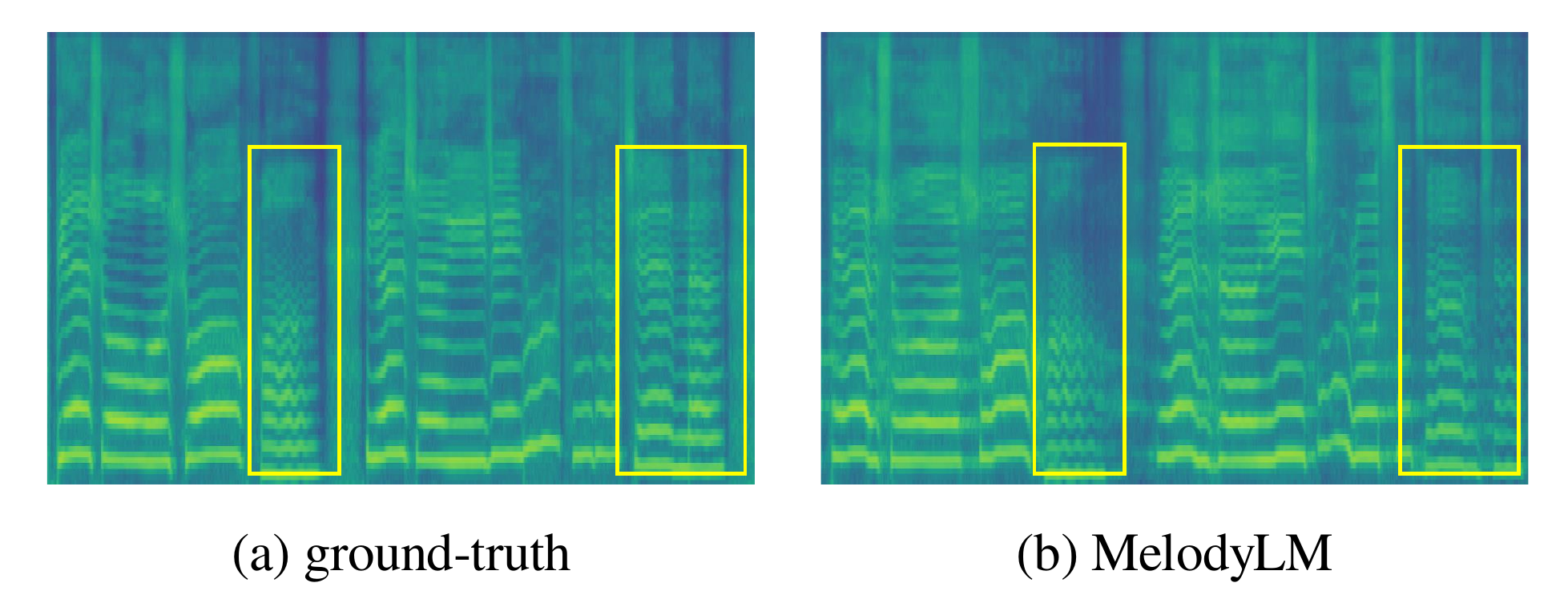}
    \caption{Visualization of the pitch and prosody modeling.}
    \label{fig:mel}
\end{wrapfigure}

However, \M~underperforms Melodist in terms of SMOS and RMSSinger in terms of MOS-Q. This may be because most of the training data of \M~are crawled and demixed using a music source separate tool \cite{ultimatevocalremovergui}, where there are unavoidable harmonics, as well as some reverb and sound effects. Voices generated by \M~could have random harmonics and reverbs, making them more unnatural. In addition, the reference vocal prompts are selected automatically and randomly, while some of them could have considerable harmonics with different timbres from the main vocal, making the model clone the wrong voice. 

\subsection{Accompaniment Generation and Song Generation }

We compare the performance of models with different degrees of control in accompaniment generation. To control the information flow, we compare the models with or without GT vocal inputs. For \M, we further control the condition by comparing vocals generated from GT MIDI or not. In general, \M~outperforms Melodist by a large margin, which may be because the hybrid conditioning has a certain advantage over the simple autoregressive generation. The direct temporal control of vocal signals provides an accurate duration modeling, resulting in a higher MEL. The latent diffusion model is also better at modeling features with complex frequency bands, such as music, as reflected in a lower FAD. However, dropping the optional inputs (MIDI and vocal) indeed reduces the overall performance, indicating the potential error accumulation of our cascaded design.

For an ultimate evaluation, we remix the generated vocals and accompaniments. The results are listed in \autoref{tab:accomp-res}, in terms of OVL-M. \M~with full GT inputs outperforms all the baselines for better audio quality, relevance to the prompts, and interaction between vocals and accompaniments. However, \M~with only text and reference inputs still achieve considerable performance. 



\subsection{Ablation}



\paragraph{Song Generation with Minimum Conditions}

We explore the unconditional generation capability of \M, to verify the claim that \M~can generate a song piece with the minimum requirement, namely, lyrics and vocal reference prompt. We drop all the optional textual prompts and infer the cascaded models. 
The subjective SVS evaluation (the objective metrics are not applicable, since FEE is meaningless here) results are listed in \autoref{tab:svs-abl}, denoted as \M~(unc.).

\begin{wraptable}{r}{10cm}
    \caption{Ablation Results of SVS.}
    \label{tab:svs-abl}
    \footnotesize
	\centering
    \resizebox{0.7\textwidth}{!}{
	\begin{tabular}{lcccc}
    \toprule
    Methods             & FEE$\downarrow$   & SMOS$\uparrow$    & MOS-P$\uparrow$   & MOS-Q$\uparrow$   \\
    \midrule
    \M~(ours)           & 0.08              & 3.81$\pm$0.12     & 3.82$\pm$0.08     & 3.76$\pm$0.10     \\
    \midrule[0.2pt]
    \M~(unc.)           & -                 & 3.79$\pm$0.11     & 3.78$\pm$0.14     & 3.74$\pm$0.09     \\
    \M~(unexpand)       & 0.24              & 3.71$\pm$0.08     & 3.61$\pm$0.15     & 3.69$\pm$0.11     \\
    \M~(e2e w/ MIDI)    & 0.35              & 3.72$\pm$0.13     & 3.56$\pm$0.09     & 3.67$\pm$0.10     \\
    \M~(e2e w/o MIDI)   & -                 & 3.61$\pm$0.15     & 3.52$\pm$0.12     & 3.61$\pm$0.11     \\
    \bottomrule 
	\end{tabular}
 }
\end{wraptable}

For accompaniment evaluation, the results are listed in \autoref{tab:accomp-res}, also denoted as \M~(unc.) (this "unc." means no prompt, while the vocal condition is still available). The results demonstrate a performance reduction in SVS tests. We believe it is because there is certain gap between the characteristic of the vocal reference and the uncontrolled melody. During the listening test, some bad cases are found if, for example, the uncontrolled MIDI sequence has a relatively high average pitch and the reference is a deep male voice. 

\paragraph{MIDI Representation}

In the original design, the output form of MIDI tokens is unexpanded (namely, two tokens at one step: pitch and duration), while the input form in the Vocal-LM is expanded for positional information. To explore the difference between representations and prepare the following one-stage generation, we change the MIDI to the unexpanded form in the Vocal-LM and record the results, denoted as "unexpand" in \autoref{tab:svs-abl}. It can be seen that numerical duration information is difficult to perceive for our model and the prosody is largely degraded.

\paragraph{One-stage Singing Voice Synthesis}

We conduct ablation experiments on the cascaded architecture in SVS. Specifically, we connect the MIDI-LM and Vocal-LM into one multi-scale language model, with a size of 420M parameters (hidden size 1536, 12 heads, 14 layers). 
To suit the one-stage modeling, we drop the expanded form of MIDI and directly concatenate the MIDI tokens and the vocal acoustic tokens after the input prompts, where the loss is applied to both MIDI and vocal tokens. The results are listed in \autoref{tab:svs-abl}, denoted as "e2e w/ MIDI". Another comparison is conducted with the same architecture, but the MIDI tokens are dropped entirely, denoted as "e2e w/o MIDI". The overall performance is reduced dramatically due to information loss. We believe that a much larger dataset could solve this problem and enable one-stage generation, but under the current limited data conditions, a multi-stage approach is the optimal solution.

\section{Conclusion}

We introduced \M, the first accompanied singing voice synthesis model that generates high-quality song pieces with fully text-controlled melodies. We demonstrated the effectiveness of connecting textual prompts and singing voices with intermediate musical features like MIDI. We achieved controllable accompaniment generation with latent diffusion models and a hybrid conditioning mechanism. Furthermore, we provided a comprehensive study of generation with minimum conditions, conditioning methods, and one-stage singing voice synthesis. Experimental results demonstrated that the proposed model achieved state-of-the-art performance in separated or joint stages. 





{
\small

\bibliographystyle{neurips_2024}
\bibliography{neurips_2024}

}

\appendix

\section{Data}  \label{appendix:data}

The details and the statistics of the datasets used in our work are listed in \autoref{tab:data}. For the web-crawled data, we perform the following pre-processing operations:
\begin{itemize}[leftmargin=*]
    \item We use \texttt{Ultimate Vocal Remover} \cite{ultimatevocalremovergui}, an open-source music source separation tool to demix all of the songs.
    \item We utilize WhisperX \cite{bain2022whisperx} to automatically transcribe the demixed vocals and acquire the sentence-level timestamps. We set the maximal sentence length to 30 seconds and the minimal 1 second. 
    \item We use the obtained sentence-level timestamps to segment the original and the demixed songs. After that, we filter the samples using Silero VAD \cite{SileroVAD}, to eliminate unvoiced clips. 
    \item We utilize the music captioning model \cite{doh2023lp} to generate pseudo captions from the segmented mixed clips, which are eventually used as the accompaniment-related prompts. 
    \item We utilize the MIDI extraction model \cite{li2024robust} to obtain unaligned MIDI sequences from the segmented demixed vocal clips.     
\end{itemize}

\section{Textual Prompts} \label{appendix:prompt}

For controllable melody generation, we construct artificial textual prompts to deliver melody-related information. Specifically: 1) we compute the key of each vocal sample from the extracted MIDI sequence using the Krumhansl-Schmuckler algorithm \cite{krumhansl2001cognitive}, along with the corresponding Pearson correlation coefficient; 2) we compute the average pitch value for each sample and qualitatively divide it into five categories (very low, low, medium, high, and very high); 3) we determine the tempo value on the original song sample using deeprhythm\footnote{https://github.com/bleugreen/deeprhythm}\cite{aarabi2019deep} and also divide it into five categories; 4) for emotion information, we use ChatGPT \cite{chatgpt} to extract emotion-related keywords from the corresponding accompaniment-related prompt (described below); and 5) we also divide the duration of each sample into four categories. Finally, we produce $N_{\text{melody}}$ textual templates for musical attributes to construct melody-related natural language prompts. This prompt provides a general direction for melody generation.

\section{Implementation Details} \label{appendix:imp}

\paragraph{Musical Features}

We use a retrained ROSVOT MIDI extractor with no word boundary predictor for MIDI transcription. Only the datasets with unavailable MIDI annotations are transcribed. Since the granularity of MIDI note events is affected by different singing styles (e.g., different levels of portamentos), we apply data augmentation by extracting three sets of MIDI sequences for each sample using different note boundary thresholds (0.8, 0.85, and 0.9), and choose the MIDI representations randomly during training. The key computed from the MIDI sequence is randomly switched to its relative (e.g., C major to A minor), and those with Pearson correlation coefficient below 0.5 are dropped, since they could be single-pitch or rap. The tempo values with confidence below 0.3 are also dropped. To avoid model confusion, when dividing categories for average pitch, tempo, and duration, we drop the tags near the decision boundaries. 

\paragraph{Textual Features}

For the web-crawled dataset, we utilize WhisperX \cite{bain2022whisperx}, an automatic speech recognition (ASR) model with timestamp prediction, to obtain the lyrics and the temporal offset of each sentence (the word-level alignment is not used because there are always mistranscribed words with zero duration). We then use \texttt{pypinyin} to convert the lyrics into pinyin for pronunciation conditioning. For computational efficiency, we limit the max text sequence of the lyrics encoder to 80 and the melody-related prompt to 50. The limit of the accompaniment prompt encoder is set to 80. We adopt different prompt encoders to compare the controllability. 

\begin{table}[]
    \caption{Datasets in different stages and the actual lengths we use.}
    \label{tab:data}
    \footnotesize
    \centering
    \setlength\tabcolsep{3pt}
    \resizebox{\textwidth}{!}{
    \begin{tabular}{lcccccc}
    \toprule
    Dataset                             & Type              & Annotation                & MIDI Stage        & Vocal Stage       & Accomp. Stage     & Length (hrs)  \\
    \midrule
    Opencpop\cite{wang2022opencpop}     & singing           & lyrics, duration, MIDI    & \ding{51}         & \ding{51}         & \ding{55}         & 5.3     \\
    M4Singer\cite{zhang2022m4singer}    & singing           & lyrics, duration, MIDI    & \ding{51}         & \ding{51}         & \ding{55}         & 29.8    \\
    OpenSinger\cite{huang2021multi}     & singing           & lyrics, duration          & \ding{51}         & \ding{51}         & \ding{55}         & 83.5      \\
    web-crawled                         & singing, accomp.  & -                         & \ding{51}         & \ding{51}         & \ding{51}         & 693.4       \\
    LP-MusicCaps-MSD\cite{doh2023lp}    & singing, accomp.  & pseudo captions           & \ding{55}         & \ding{55}         & \ding{51}         & 306.2       \\
    \bottomrule 
	\end{tabular}
    }

\end{table}

\paragraph{Audio Tokenization and Reconstruction}

We utilize a SoundStream audio tokenizer for 16 kHz monophonic waveforms with a hop size of 320, so that the frame rate is 50 Hz. The embeddings are quantized with $N_q=8$ codebooks with a size of 1024. To recover the vocal waveforms from 3 codebooks in our design, we train a unit-based HiFi-GAN with an upsampling layer to reconstruct 24 kHz waveforms. For accompaniment generation, the 3-layer 1D VAE with a latent dimension of 20 downsamples the Mel-spectrograms with a ratio of 2, where the spectrograms are computed with 80 frequency bins and a hop size of 320 from 24 kHz waveforms. A corresponding Mel-based HiFi-GAN for 24 kHz accompaniments is also pre-trained. 

\paragraph{Language Modeling}

For the MIDI-LM, the global transformer is 16-layer and 12-head with a hidden dimension of 768 and 110M parameters, while the local transformer is similar but with 6 layers and 8 heads and 40M parameters. For the Vocal-LM, the architecture is shared, but the global transformer is 20-layer and 16-head with a hidden dimension of 1152 and 320M parameters, while the local transformer has 6 layers and 8 heads with 100M parameters. We train both models using the Adam optimizer with $\beta_1 = 0.9$, $\beta_2 = 0.98, \epsilon = 10^{-8}$, and a base learning rate of $5 \times 10^{-4}$. The MIDI-LM is trained for 50K steps using 4 NVIDIA V100 GPUs with a batch size of 1400, and the Vocal-LM is trained for 100K steps using 6 GPUs with 12K max tokens. 

\paragraph{Latent Diffusion Model}

The denoiser is a 4-layer FFT with a hidden dimension of 576, a convolutional fead-forward dimension of 2304, and a kernel size of 9, resulting in 160M parameters. The vocal features are upsampled by a factor of 1.5 before being fed into the model. The diffusion model is trained on 4 NVIDIA V100 GPUs with 80K steps and a batch size of 240 samples. An AdamW optimizer is adopted with a base learning rate of $3 \times 10^{-6}$. 

\section{Training Procedure} \label{appendix:train}

The three stages are trained separately: (i) For stage 0, we use the last hidden vectors from a text encoder, BERT (base) \cite{devlin2018bert}, to provide semantic information that preserves temporal length, since vocal melody composition always takes the number of words and the potential emotion-related descriptions in the lyrics into account. Another text encoder, FLAN-T5 \cite{chung2024scaling}, if not otherwise stated, is utilized to encode the prompt for better controllability. The duration values of MIDI notes are translated into positional offsets for robust decoding.  (ii) For stage 1, we directly tokenize the pinyin of preprocessed lyrics for language modeling, since the voice synthesis stage only requires pronunciation information. The reference acoustic tokens are extracted from the vocal clips with the same singer and the same song. Both the MIDI and vocal LM are trained with the conventional negative log-likelihood loss, where only the MIDI tokens and the vocal tokens are considered in the loss. (iii) We combine the melody-related prompts in stage 0 and the accompaniment-related prompts together before the conditional denoising. The dropout rates of each condition separately and jointly are all 0.1, resulting in a final dropout rate of 0.19. A 1D VAE model is pre-trained to reconstruct Mel-spectrograms from the latent features.




\section{Limitation and Potential Negative Impact}

\paragraph{Limitations}

The proposed model has three stages, relying on multiple infrastructures like vocoders, VAE, etc., resulting in a cumbersome training and inference procedure. Also, the corpus only contains Mandarin pop songs, lacking diversity.

\paragraph{Potential Risks} 

Large-scale generative models present ethical challenges. Misuse of the proposed model may lead to copyright issues. Proper constraints are needed to guarantee people who use our code or pre-trained models will not use the model in illegal cases.


\newpage

\end{document}